\documentstyle[twocolumn,prb,aps,epsfig]{revtex}

\begin{document}
\title{Quantum computing of molecular magnet Mn$_{12}$}
\author{Bin Zhou$^{1,2,3}$, Ruibao Tao$^{1\text{,}\dagger }$, Shun-Qing Shen$^{4}$,
and Jiu-Qing Liang$^{5}$}
\address{1. Department of Physics, Fudan University, Shanghai 200437, China\\
2. State Key Laboratory of Magnetism, Institute of Physics , Chinese Academy%
\\
of Sciences, P.O. Box 603-12, Beijing 100080, China\\
3. Department of Physics, Hubei University, Wuhan 430062, China\\
4. Department of Physics, The University of Hong Kong, Hong Kong, China\\
5. Institute of Theoretical Physics, Shanxi University, Taiyuan 030006,China}
\date{\today }

\twocolumn[\hsize\textwidth\columnwidth\hsize\csname@twocolumnfalse\endcsname

\maketitle

\begin{abstract}
\renewcommand{\thefootnote}{\fnsymbol{footnote}} 

Quantum computation in molecular magnets is studied by solving the
time-dependent Schr\"{o}dinger equation numerically. Following Leuenberger
and Loss (Nature (London) 410, 789(2001)), an external oscillating magnetic
field is applied to populate and manipulate the spin coherent states in
molecular magnet Mn$_{12}$. The conditions to realize parallel recording and
reading data bases of Grover algorithsm in molecular magnets are discussed
in details. It is found that an accurate duration time of magnetic pulse as
well as the amplitudes are required to design the device of quantum
computing.
\end{abstract}


]


Quantum phase was proposed to store information in connection with a new
class of computational algorithms based on the rules of quantum mechanics
rather than classical physics.\cite{Deutsch} One example is the database
search problem proposed by Grover.\cite{Grover(1),Grover(2)} Differing from
other quantum algorithms, the superposition of single-particle quantum
states is sufficient for Grover's algorithm. The Grover's algorithm was
successfully implemented using Rydberg atoms.\cite{Ahn} Recently,
Leuenberger and Loss theoretically proposed that molecular magnets Mn$_{12}$
can be used to realize the Grover's algorithm by utilizing multifrequency
coherent magnetic radiation.\cite{Leuenberger} In the S-matrix and
time-dependent high-order perturbation theory, they showed that it was
possible to populate and manipulate spin coherent excited states by applying
a single pulse of weak oscillating transverse magnetic field with an
appropriate number of matching frequencies, and \ the coherent state can be
applied for storing a multi-bites information. In this paper the population
and manipulation of a spin coherent excited state in an external oscillating
magnetic field is studied by solving the time-dependent Schr\"{o}dinger
equations numerically. The conditions for the field to implement the
Grover's algorithm are discussed in details. We find that an accurate
duration time of magnetic pulse as well as the oscillating frequencies and
amplitudes are required to design the device of quantum computing in the
molecular magnets Mn$_{12}.$

Following Leuenberger and Loss, we consider a molecular magnet with spin $S$ 
$(>1/2)$ in presence of a weak oscillating transverse magnetic field. The
Hamiltonian for the system reads 
\begin{equation}
H=H_{\text{spin}}+V_{\text{low}}(t)+V_{\text{high}}(t),
\end{equation}
where 
\begin{equation}
H_{\text{spin}}=-AS_{z}^{2}-BS_{z}^{4}+g\mu _{B}\delta H_{z}S_{z},
\label{H0}
\end{equation}

\begin{equation}
V_{\text{low}}(t)=g\mu _BH_0(t)\cos (\omega _0t)S_z,
\end{equation}

\begin{eqnarray}
V_{\text{high}}(t) &=&\sum_{m=m_{0}}^{s-1}\frac{g\mu _{B}H_{m}(t)}{2} 
\nonumber \\
&&\times \left[ e^{i(\omega _{m}t+\Phi _{m})}S_{+}+e^{-i(\omega _{m}t+\Phi
_{m})}S_{-}\right] .
\end{eqnarray}
$H_{\text{spin}}$ is a single spin Hamiltonian for molecular magnet Mn$_{12}.
$\cite{Barra} $\left| m\right\rangle $ are the simultaneous eigenstates of $%
H_{\text{spin}}$ and the z-component spin operator ${\bf S}_{z}$ with the
energy $\varepsilon _{m}$ and the moment $m\hbar $. $V_{\text{low}}$ and $V_{%
\text{high}}$ are the Zeeman terms to describe the coupling between the
external magnetic fields and the spin S. The low frequency is applied along
the easy axis, and $V_{\text{low}}$ supplies the necessary energy for the
resonance condition. The high frequency transverse fields are introduced to
induce the transition from the initial state $\left| s\right\rangle $ to
virtual states $\left| m\right\rangle ,$ $m=m_{0},\cdots ,s-1,$ and the
frequencies $\omega _{m}$, $m=m_{0},\cdots ,s-2,$ mismatch the level
separation by $\omega _{0}$, that is , $\hbar \omega _{m}=\varepsilon
_{m}-\varepsilon _{m+1}+\hbar \omega _{0},$ and $\omega _{s-1}=$ $%
\varepsilon _{s}-\varepsilon _{s+1}+(s-m_{0}-1)\hbar \omega _{0}.$ The
phases $\Phi _{m}=\sum_{k=s-1}^{m+1}\Phi _{k}+\varphi _{m\text{ }}$($\varphi
_{m\text{ }}$ is the relative phase) and 
\[
H_{m}(t)=\left\{ 
\begin{array}{l}
H_{m},\text{ if }t\in (-\frac{T}{2},\frac{T}{2})\text{ for }m=0\text{ and }%
m\geq m_{0}, \\ 
0,\text{ otherwise.}
\end{array}
\right. 
\]
The time-dependent Schr\"{o}dinger equation reads 
\begin{equation}
i\hbar \frac{\partial }{\partial t}\left| \Psi (t)\right\rangle =H\left|
\Psi (t)\right\rangle ,  \label{SE}
\end{equation}
where 
\begin{equation}
\left| \Psi (t)\right\rangle =\sum_{m=m_{0}}^{s}a_{m}(t)e^{-i\varepsilon
_{m}t/\hbar }\left| m\right\rangle   \label{6}
\end{equation}
The eigenstates of $m<m_{0}$ have been neglected. The molecular magnets Mn$%
_{12}$ behaves like single spin, and has a ground state with spin $s=10$.

The Grover scheme becomes adapted to describe the quantum computational
read-in and decoding of the quantum data register ${\bf a=(}%
a_{s},a_{s-1},...,a_{m_{0}}).$ Firstly, as a simplest example, let us
discuss the case of $m_{0}=9.$ At the first stage of the so called
``read-in'', the initial values of $\{a_{m}^{(0)}:m=10,9\}$ are set to be $%
\{a_{10}^{\left( 0\right) }=1$, $a_{9}^{\left( 0\right) }=0\}.$ After
irradiating the molecular magnets Mn$_{12}$ with a coherent magnetic pulse
with the relative phase $\varphi _{9}^{(1)}$ in the duration time $T$, one
has the quantum data register ${\bf a=(}a_{10}^{(1)},a_{9}^{(1)})$ with 
\begin{equation}
a_{10}^{(1)}=\cos \left( \frac{g\mu _{B}H_{9}}{2\hbar }\langle
9|S_{-}|10\rangle T\right) ,  \label{7}
\end{equation}
\begin{equation}
a_{9}^{\left( 1\right) }=-ie^{-i\varphi _{9}^{(1)}}\sin \left( \frac{g\mu
_{B}H_{9}}{2\hbar }\langle 9|S_{-}|10\rangle T\right) .  \label{8}
\end{equation}
At the second stage of the so called ``decoding'', the solutions of (\ref{7}%
) and (\ref{8}) become the initial values of $\{a_{m}\}$ in Eq.(\ref{SE}),
and the relative phase is set to be $\varphi _{9}^{(2)}=0.$ At the end of
decoding, one obtains

\begin{equation}
a_{9}^{(2)}=\left\{ 
\begin{array}{l}
-i\sin \left( 2\frac{g\mu _{B}H_{9}}{2\hbar }\langle 9|S_{-}|10\rangle
T\right) ,\text{ \quad for }\varphi _{9}^{(1)}=0 \\ 
0,\text{ \qquad \qquad \qquad \qquad \qquad \qquad for }\varphi
_{9}^{(1)}=\pi
\end{array}
\right.  \label{simplest}
\end{equation}
If the condition of $g\mu _{B}H_{9}\langle 9|S_{-}|10\rangle T/2\hbar \ll 1$
is satisfied, the result in Eq. (\ref{simplest}) is in a good consistence
with Ref.[5]. For a longer duration time $T\gg \left[ g\mu _{B}H_{9}\langle
9|S_{-}|10\rangle /2\hbar \right] ^{-1},$the condition fails to implement
the Grover algorithm.

For $m_{0}<9$, it is difficult to solve Eq.(\ref{SE}) analytically. However,
as it is a single particle problem only relating to a finite number of
states, it is possible for us to solve the differential equations
numerically instead of evaluating the S-matrix perturbatively as Leuenberger
and Loss did. The results of explicit numerical calculation should be more
reliable than other approximate results. For illustration, we focus on the
case $m_{0}=5.$ The parameters used here are the same as the ones in
Ref.[5]: $\omega _{0}=5\times 10^{7}s^{-1}$, $T=10^{-7}s$, $H_{0}=H_{9}=2G$, 
$H_{8}/H_{0}=-0.04,$ $H_{7}/H_{0}=-0.25,$ $H_{6}/H_{0}=-0.61,$ and $%
H_{5}/H_{0}=-1.12$. For the molecular magnets Mn$_{12}$, $A$ and $B$ in Eq.(%
\ref{H0}) are $0.56K$ and $1.11\times 10^{-3}K,$ respectively.\cite{Barra}
According to the Ref.[5], the relative phases are $\varphi
_{9}^{(1)}=\varphi _{8}^{(1)}=\varphi _{7}^{(1)}=0$ and $\varphi
_{6}^{(1)}=\varphi _{5}^{(1)}=\pi $ for encoding the number $%
13_{10}=1101_{2}.$ Now, we solve six first-order differential equations for $%
\{a_{m}(t)\}$ in Eq. (\ref{6}). The initial values of the $\{a_{m}\}$ are
set to be $a_{10}^{(0)}=1$ and $a_{m}^{(0)}=0$ for $m\neq 10$. After a
duration time T, we have ${\bf a}^{(1)}{\bf =(}%
a_{10}^{(1)},a_{9}^{(1)},a_{8}^{(1)},a_{7}^{(1)},a_{6}^{(1)},a_{5}^{(1)}).$
At the second stage for decoding the number, ${\bf a}^{(1)}$ becomes the
initial values of Eq.(\ref{SE}) and the relative phases should be set by $%
\varphi _{9}^{(2)}=\varphi _{7}^{(2)}=\varphi _{5}^{(2)}=0$ and $\varphi
_{8}^{(2)}=\varphi _{6}^{(2)}=\pi $ . Once again, the six first-order
differential equations with the new initial conditions are solved
numerically. At the end of decoding, we obtain ${\bf a}^{(2)}{\bf =(}%
a_{10}^{(2)},a_{9}^{(2)},a_{8}^{(2)},a_{7}^{(2)},a_{6}^{(2)},a_{5}^{(2)}).$
If Grover's algorithm is implemented successfully, according to Leuenberger
and Loss, one should have $\left| a_{10}^{(2)}\right| \approx 1,\left|
a_{9}^{(2)}\right| \approx 2\left| a_{9}^{(1)}\right| \approx 2\eta ,$ $%
\left| a_{8}^{(2)}\right| \approx 0,$ $\left| a_{7}^{(2)}\right| \approx
2\left| a_{7}^{(1)}\right| \approx 2\eta ,$ $\left| a_{6}^{(2)}\right|
\approx 2\left| a_{6}^{(1)}\right| \approx 2\eta ,$ $\left|
a_{5}^{(2)}\right| \approx 0$ and $\eta \ll 1.$ Instead, our numerical
calculation gives 
\begin{eqnarray*}
\left| a_{10}^{(1)}\right|  &\approx &0.95,\text{ }\left| a_{9}^{(1)}\right|
\approx 0.31,\text{ }\left| a_{8}^{(1)}\right| \approx 0.04,\text{ } \\
\left| a_{7}^{(1)}\right|  &\approx &0.01,\text{ }\left| a_{6}^{(1)}\right|
\approx 0.02,\text{ }\left| a_{5}^{(1)}\right| \approx 0.01
\end{eqnarray*}
and 
\begin{eqnarray*}
\left| a_{10}^{(2)}\right|  &\approx &0.82,\text{ }\left| a_{9}^{(2)}\right|
\approx 0.59,\text{ }\left| a_{8}^{(2)}\right| \approx 0.08, \\
\left| a_{7}^{(2)}\right|  &\approx &0.06,\text{ }\left| a_{6}^{(2)}\right|
\approx 0.01,\text{ }\left| a_{5}^{(2)}\right| \approx 0.01.
\end{eqnarray*}
The values of $\{a_{m}^{(1)},a_{m}^{(2)}\}$ are complex numbers (not real)
and we only present their absolute values here. Therefore, the numerical
solution shows that the encoding number likes $0001_{2},$ but is not $%
1101_{2}$ as expected by Leuenberger and Loss at the specific values of
applied field and duration time $T$ . However, $\{a_{m}^{(1)},a_{m}^{(2)}\}$
depend on the duration time $T$, and one may be still able to reach to the
encoding number $1101_{2}$ by a specific duration time $T$ of pulse.

To clarify the feasibility of the method proposed by Leuenberger and Loss,
we study a simple, but nontrivial case of $m_{0}=8$ as a representative
example. The key approximations in Leuenberger and Loss's method are that $%
S_{m,s}^{(j)}=0$ for $j<n(=s-m_{0})$. The transitions from $|k\rangle $ to $%
|m\rangle $, i.e. $S_{m,k}^{(j)}$ $(m<k<s)$ and all higher order amplitudes $%
S_{m,s}^{(j)}$ are neglected$.$ In the case of $m_{0}=8$, the first order
perturbation are required to be zero: $S_{9,10}^{(1)}=S_{8,9}^{(1)}=0,$ or
at least $\left| S_{9,10}^{(1)}\right| \ll \left| S_{9,10}^{(2)}\right|
(\simeq \left| S_{8,10}^{(2)}\right| )$ and $\left| S_{8,9}^{(1)}\right| \ll
\left| S_{9,10}^{(2)}\right| (\simeq \left| S_{8,10}^{(2)}\right| ).$ In
fact, the quantum amplitudes of the transitions by the perturbative $S-$%
matrix formula are given by

\begin{equation}
S_{9,10}^{(2)}=\frac{2\pi }i\left( \frac{g\mu _B}{2\hbar }\right) ^2\frac{%
H_0H_9e^{i\Phi _9}}{\omega _0}\langle 9|S_{-}|10\rangle \frac T{2\pi }
\label{S9-10-1}
\end{equation}

\begin{equation}
S_{8,10}^{(2)}=\frac{2\pi }i\left( \frac{g\mu _B}{2\hbar }\right) ^2\frac{%
H_8H_9e^{i\Phi _8}e^{i\Phi _9}}{\omega _0}\langle 8|S_{-}|9\rangle \langle
9|S_{-}|10\rangle \frac T{2\pi }  \label{S8-10-1}
\end{equation}

\begin{equation}
S_{9,10}^{(1)}=\frac 1i\frac{g\mu _BH_9e^{i\Phi _9}}{2\hbar }\langle
9|S_{-}|10\rangle \frac{\sin (\omega _0T/2)}{\omega _0/2}  \label{S9-10}
\end{equation}

\begin{equation}
S_{8,9}^{(1)}=\frac{1}{i}\frac{g\mu _{B}H_{8}e^{i\Phi _{8}}}{2\hbar }\langle
8|S_{-}|9\rangle \frac{\sin (\omega _{0}T/2)}{\omega _{0}/2}  \label{S8-9}
\end{equation}
where $\Phi _{9}=\varphi _{9}$ and $\Phi _{8}=\Phi _{9}+\varphi _{8}$ ($%
\varphi _{9}$ and $\varphi _{8}$ are the relative phase). The first two
terms $\{S_{9,10}^{(2)},S_{8,10}^{(2)}\}$ result from the second
perturbation contributions with energy conservation. From the condition $%
\left| S_{9,10}^{(2)}\right| \simeq \left| S_{8,10}^{(2)}\right| ,$ one can
deduce the field amplitude $H_{8}/H_{0}=0.16.$ However, the lower order
terms $S_{9,10}^{(1)}$ and $S_{8,9}^{(1)}$ , in general, are not exactly
equal to zero even if they do not satisfy the energy conservation. They can
be zero only if $\omega _{0}T=2\pi l$ where $l$ is an integer. Due to these
terms $S_{9,10}^{(1)}$ and $S_{8,9}^{(1)}$ , the amplitudes of the $a_{8}(t)$
and $a_{9}(t)$ in the state $\left| \Psi (t)\right\rangle $ are obviously
oscillating with varying $T$. We have shown them in Fig. (1).

Now, let us assume the parameters: $T=1.0\times 10^{-9}$s, $H_{0}=200G$, $%
H_{9}=20G$ and $\omega _{0}=4\pi \times 10^{9}$s$^{-1}$. It gives $\left|
S_{8,10}^{(2)}\right| \simeq 0.11.$ Following Leuenberger and Loss, we set $%
\varphi _{9}^{(2)}=0$ and $\varphi _{8}^{(2)}=0$. Then, the recording number 
$00_{2}$ corresponds to $\{\varphi $ $_{9}^{(1)}=\pi $, $\varphi
_{8}^{(1)}=\pi \},$ $01_{2}$ to $\{\varphi _{9}^{(1)}=0$, $\varphi
_{8}^{(1)}=\pi \},$ $10_{2}$ to $\{\varphi _{9}^{(1)}=\pi $, $\varphi
_{8}^{(1)}=0\},$ and $11_{2}$ to $\{\varphi _{9}^{(1)}=0$, $\varphi
_{8}^{(1)}=0\}.$ The results by our numerical calculation are shown in the
Table 1 (where $m=9,8$ represent the binary digits $2^{0}$ and $2^{1},$
respectively).

Table 1: The numerical results for the case of $m_{0}=8$.

\[
\begin{tabular}{|c|c|c|c|c|c|c|}
\hline
decoded number & $\left| a_{10}^{(1)}\right| $ & $\left| a_{9}^{(1)}\right| $
& $\left| a_{8}^{(1)}\right| $ & $\left| a_{10}^{(2)}\right| $ & $\left|
a_{9}^{(2)}\right| $ & $\left| a_{8}^{(2)}\right| $ \\ \hline
$00_{2}$ & $0.99$ & $0.13$ & $0.10$ & $1.00$ & $0.07$ & $0.01$ \\ \hline
$01_{2}$ & $0.99$ & $0.13$ & $0.10$ & $0.96$ & $0.26$ & $0.04$ \\ \hline
$10_{2}$ & $0.98$ & $0.14$ & $0.10$ & $0.98$ & $0.04$ & $0.20$ \\ \hline
$11_{2}$ & $0.98$ & $0.14$ & $0.10$ & $0.94$ & $0.28$ & $0.19$ \\ \hline
\end{tabular}
\]
\newline
Using the above parameters, the proposal by Leuenberger and Loss can be
implemented. However, the values $\{|a_{10}^{(2)}|,$ $|a_{9}^{(2)}|,$ $%
|a_{8}^{(2)}|\}$ are not constants and will be changed with the duration
time $T$ of pulse. In a certain range, $\left| S_{9,10}^{(1)}\right| $ or $%
\left| S_{8,9}^{(1)}\right| $ will have the close magnitude to $\left|
S_{m,s}^{(n)}\right| $, which can result in the deformation of the
recording, or even in the destruction. In Fig.1 we plot amplitudes of $%
\left| a_{9}^{(2)}\right| $ and $\left| a_{8}^{(2)}\right| $ with varying $T$
(for $\varphi _{9}^{(1)}=0$ and $\varphi _{8}^{(1)}=\pi $) which corresponds
to $01_{2}.$ The amplitudes $\left| a_{9}^{(2)}\right| \sim 0.26$ and $%
\left| a_{8}^{(2)}\right| \sim 0.04$ for a duration time $T=10^{-9}s.$ The
value $0.26$ can be considered as a digit number $1$ and $\ 0.04$ as $0.$
The values $\left| a_{9}^{(2)}\right| $ and $\left| a_{8}^{(2)}\right| $ are
varied with the pulse duration $T.$ The first maximum of $\left|
a_{8}^{(2)}\right| $ $\sim 0.107$ at $T\sim 1.16$ $\times 10^{-9}s,$ the
second maximum $\left| a_{8}^{(2)}\right| \sim 0.22$ at $T\sim 1.66$ $\times
10^{-9}s$ which becomes larger than the first minium of $\left|
a_{9}^{(2)}\right| \sim 0.19$ at $T\sim 1.37$ $\times 10^{-9}s.$ Therefore,
we conclude that a rather definite duration of pulse should be chosen for a
stable and clear quantum recording.

One can straightforwardly extend the above discussion to the case of $%
m_{0}=5,$ but it does not change the qualitative conclusion we drew from the
case of $m_{0}=8$. The more quantum states are used to store information,
the higher precision are required to design the magnetic pulse. It is clear
that the lower order perturbation series, $j<n$ , of the S-matrix do not
vanish for general duration $T$ even the relative transitions do not satisfy
energy conservation. Therefore, to implement the encoding and decoding in
the molecular magnets by applying an external oscillating magnetic field,
one should be very careful to choose the appropriate time duration $T$ as
well as the matching frequencies and amplitudes of the field.

\begin{figure}[tbp]
\epsfig{file=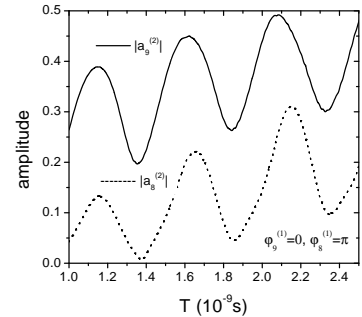, width=9.5cm}
\caption{Amplitudes $\left| a_{9}^{(2)}\right| $ and $\left|
a_{8}^{(2)}\right| $ as a function of pulse duration $T$ (for $\protect%
\varphi _{9}^{(1)}=0$ and $\protect\varphi _{8}^{(1)}=\protect\pi $)}
\end{figure}

The work is supported by the National Natural Science Foundation of China,
Shanghai Research Center of Applied Physics, Institute of Physics of Chinese
Academy of Sciences, and a CRCG grant of the University of Hong Kong.

$^{\dagger }$E-mail: rbtao@fudan.ac.cn

\end{document}